# The POOL Data Storage, Cache and Conversion Mechanism

D. Düllmann, M. Frank, G. Govi, I. Papadopoulos, S. Roiser
*European Laboratory for Particle Physics (CERN), Genève, Switzerland*

The POOL data storage mechanism is intended to satisfy the needs of the LHC experiments to store and analyze the data from the detector response of particle collisions at the LHC proton-proton collider. Both the data rate and the data volumes will largely differ from the past experience. The POOL data storage mechanism is intended to be able to cope with the experiment's requirements applying a flexible multi technology data persistency mechanism. The developed technology independent approach is flexible enough to adopt new technologies, take advantage of existing schema evolution mechanisms and allows users to access data in a technology independent way. The framework consists of several components, which can be individually adopted and integrated into existing experiment frameworks.

## 1. INTRODUCTION

The goal of the POOL project is to be able to store various types of data, which can be categorized according to their nature and role during data processing activities at the LHC experiments being prepared at the LHC collider at CERN [1]. The POOL data storage and data access mechanism is part of the POOL data persistency framework [2] and allows physicists of the LHC experiments to share the data produced by particle collisions at the collider experiments and later refined and reprocessed in worldwide distributed computing facilities. One of the most important design features of an experiment software framework is the way data persistency issues are handled. The reasons for the approach taken in POOL are described in the following sections.

## 2. A TECHNOLOGY NEUTRAL SOLUTION

Persistency support means storage and retrieval of objects currently defined in C++ across process boundaries. This support ideally is realized without intrusion into experiments' current event models, and without requiring run-time or link-time dependence between those models and the experiment's persistency technology choices.

These considerations have led us to conclude that our software architecture should support in a transparent way the use of different persistency solutions for managing the various types of data that must be treated in our data processing applications. First the volumes for the different data categories vary by many orders of magnitude. The event data representing the detector response from particle collisions from the different processing stages (raw data, reconstructed data and summary data) account for several PB/year. Data describing the state of the detector while recoding the events typically demand some TB/year. Other small amounts of data such as configuration and bookkeeping data will require only several GB per year.

Second, the different access patterns are typical for these different data stores e.g. write-once, read-many for event data, read and write many for other data, sequential access, random access, etc.

For these reasons we believe that a single persistency technology may not be optimal in all cases.

The POOL software architecture has been designed such that the best-adapted technology can be used transparently for each category of data. Data are solely accessed through the transient data cache, which exposes all required functionality to store and retrieve data. To manage the huge amount of event data, in addition to simple storage and retrieval, placement control to steer the physical data clustering is possible.

This approach, partially inspired by the work of other experiments [3,4] will allow evolving smoothly with time to more appropriate solutions as they appear in the future.

In the following the data cache mechanism, the data conversion and storage mechanism of the POOL persistency framework are described.

### 2.1. The Transient Data Cache

The goal of the POOL architecture is to impose as few restrictions as possible on the object to be made persistent such as common base classes etc. A physics algorithm can deposit objects into the transient data cache, which should be made persistent. The data cache is managed by a dedicated service, the *data service*.

Data services may exist in several instances e.g. depending on the nature and the lifetime of the objects each service manages. These groups of objects may be handled differently e.g. by applying an experiment policy:

- Event data, which get flushed after the processing of one single event
- Detector data and calibration data
- Statistical data, such as histograms.

The main programmatic interface to the data service is implemented using a smart pointer approach through so-called *object references*. Figure 1 shows how clients can access the different data services using this reference mechanism. The references also ensure type save data access.

Any object in the persistent world is identified by a *token*. This token describes the location the object in its persistent state. The token also allows distinguishing the object type in a platform independent manner. For the data-cache-service both representations, the transient object and the token are equivalent: the presence of a token allows to load the object from the persistent storage as illustrated in Figure 2, and on the other hand registering an object for persistency results in a token, which in turn can be used to uniquely identify and load the object. This identity can also be used to persist object relationships.

Each entry in the transient data cache may contain data members, which are primitives, aggregated objects or object associations to other objects. Object associations have been implemented using reference links in which the node does not acquire ownership of the referenced item. The ownership of any object belongs to the data cache, which through reference counting determines whether an object is still accessed by clients or can be dropped.

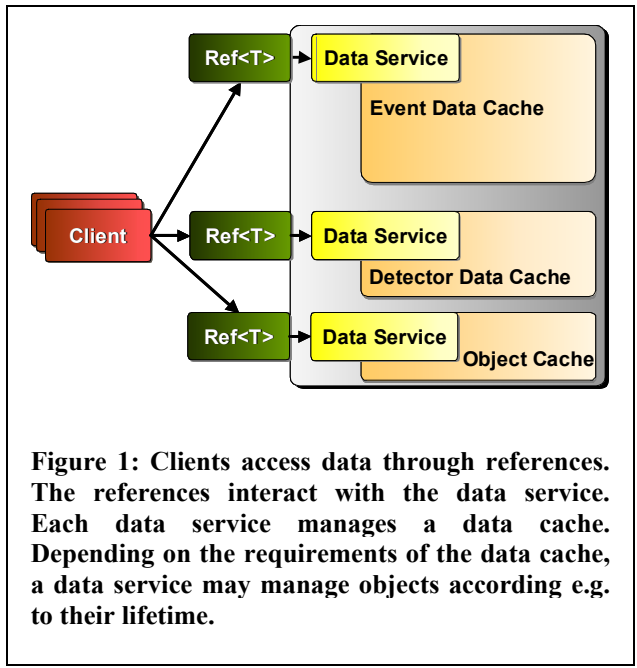

**Figure 1: Clients access data through references. The references interact with the data service. Each data service manages a data cache. Depending on the requirements of the data cache, a data service may manage objects according e.g. to their lifetime.**

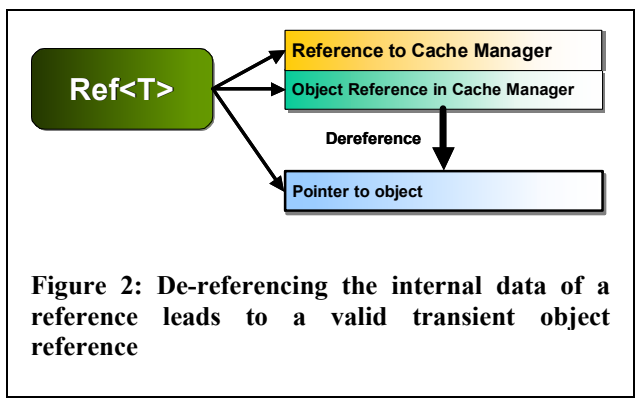

**Figure 2: De-referencing the internal data of a reference leads to a valid transient object reference**

Attention was given to the design of the data service in order to keep it as an independent component, which can easily be replaced by existing cache components of the data processing frameworks in various experiments [5,6]. Hence, although the data service knows about the existence of tokens, it does not interpret the token, but only passes the token to the data conversion mechanism. The data content of a token is explained in section 2.3.

## 2.2. The Data Conversion Mechanism

There are several options for maintaining both data representations. One is to describe the user-data types within the persistent storage framework (meta-data) and have utilities able to automatically create both representations using this meta-data. This approach is elegant under the assumption that the physical object layout does not change between different platforms, compilers etc.

Another possibility is to use generated code in order to describe the layout of the transient object, and this is the approach chosen in POOL. A technology free description of the transient object layout was used to access the object's data binding. This component, the *Data Dictionary* [7] was used to program the persistent backend with the internal layout of the object data.

In the event that the persistent backend allows describing objects as in ROOT I/O [8] (see Figure 3), this mechanism can benefit from such features. Using this technique we do not only benefit from the ROOT schema evolution mechanism, but also when analyzing data interactively because the nature of the objects is preserved.

Non-trivial gateways, which typically do have the flexibility of describing objects, depend strongly on the persistent technology. It will hence in the future be necessary to implement such gateways also for other technologies. On the other hand, simple gateways, which for example only map tabular data to homomorphous objects, can be implemented generically.

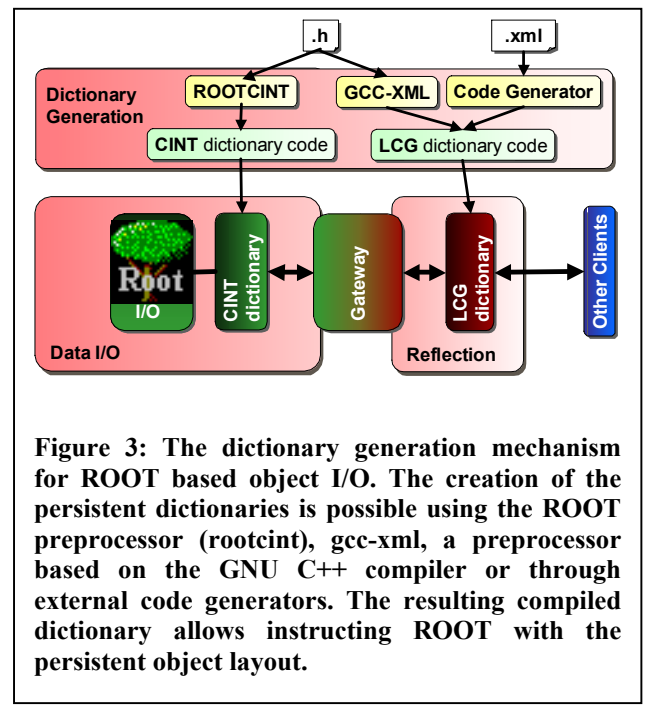

**Figure 3: The dictionary generation mechanism for ROOT based object I/O. The creation of the persistent dictionaries is possible using the ROOT preprocessor (rootcint), gcc-xml, a preprocessor based on the GNU C++ compiler or through external code generators. The resulting compiled dictionary allows instructing ROOT with the persistent object layout.**

Attention was given to allow the persistent and the transient representation of a given object to be identical. This approach avoids the necessity to reformat objects before clients can deliver them. In the event a non-standard transformation of the persistent data is required to retrieve a given transient shape of an object, a *transformation callback* can perform these complicated operations, such as the combination of many small transient objects into a single object in order to minimize overhead in storage space and I/O. When converted to the transient representation, the persistent representation is expanded to the individual objects.

Another example is the regrouping of information spread over several persistent items into a new object. Such flexibility however requires specifically written code.

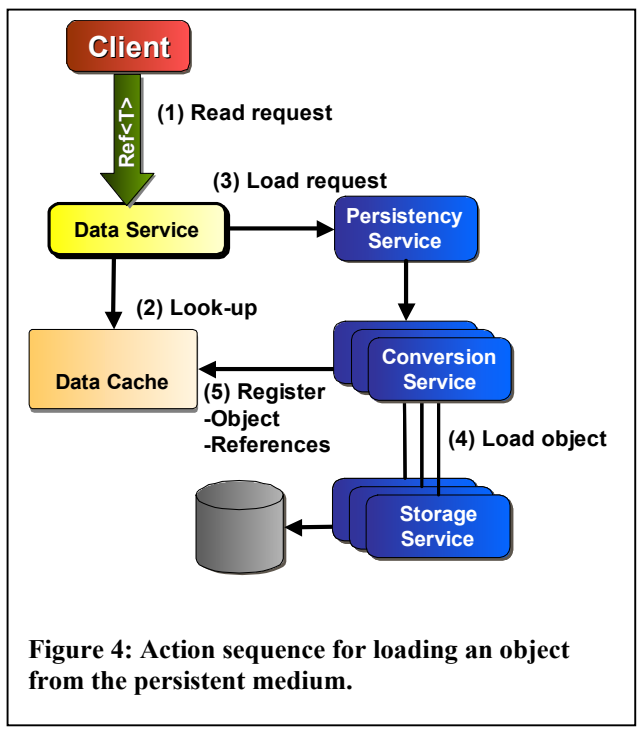

**Figure 4: Action sequence for loading an object from the persistent medium.**

Every request for an object from the data service invokes the sequence shown in Figure 4:
1. The client initiates a request to access an object.
2. The data service searches the data store for the requested object. If the object exists, a reference is returned and the sequence ends. Any object requested is identified by its token.
3. Otherwise the request is forwarded to the persistency service. The persistency service dispatches the request to the appropriate conversion service capable of handling the specified storage technology.
4. The functionality of the conversion service is split in two, where a generic conversion service handles all technology independent aspects, whereas a technology specific component, the *storage service* handles the aspects, which differ between technologies.
    o Tokens only specify a given database by its *file identifier* (FID). In a first step, the conversion service retrieve from the *file catalog* component [9] the path to the corresponding physical file name. The catalog component however is not limited to only perform the lookup, but could also invoke more complex actions like file replications etc.
    o The conversion service determines the transformation from the persistent object to the requested transient object. By default this transformation is trivial and the persistent object shape is identical to the transient object shape delivered to the client.
    o The storage service instructs the persistent technology about the desired object shape and retrieves the object. At this stage the underlying storage technology and the object description derived from the dictionary interact.
5. Before the client may use the object, any token representing a reference from the currently loaded object to other objects must be registered with the data cache to allow loading these objects on demand.

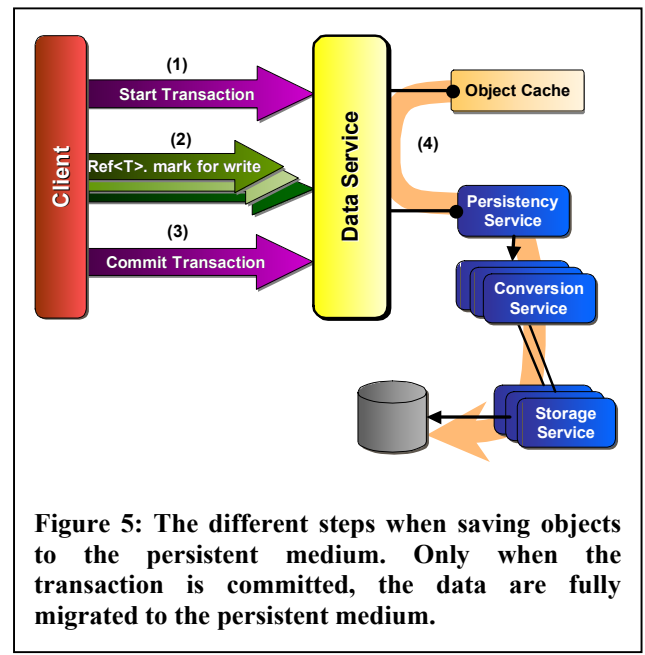

**Figure 5: The different steps when saving objects to the persistent medium. Only when the transaction is committed, the data are fully migrated to the persistent medium.**

When making objects persistent, the calling sequence is as follows (see Figure 5):
1. The client starts a data transaction. A transaction is limited to the context of one logical database. Several transactions may be open at any time as long as they do not refer to the same database.

2. The client declares iteratively one or several object to be marked for migration to the persistent storage. On every request a placeholder for the persistent object is allocated. The operation returns a unique object token, which identifies the object. The total amount of objects, which can be marked for write, depends solely on the available memory.
3. The client commits the transaction.
4. When the transaction is committed, all marked objects, which still reside in the data cache, are migrated to the persistent medium. During this process, all references in a migrating object are converted to tokens and stored as well.

## 2.3. A Generic Persistency Model

Traditionally HEP data was accessed through sequential files. The file was organized in logical records representing one event was partitioned into structures called banks.

The drawback of this sequential file organization is the difficulty to access banks from previous processing steps, and for example to re-run a reconstruction algorithm while analyzing event summary data. Such a behavior however does not result from the use of files, but rather from the inability of existing persistency mechanisms to address individual data items within other files and to read them without scanning the file.

ROOT I/O, relational (RDBMS) and object database (ODBMS) technologies allow this type of random access. Storing primitive properties of an object with these technologies is simple, but it is difficult to store references to other objects, as these pointers are only valid in the current address space and need special care.

Commercial databases solve this problem by replacing the reference with an object identifier (OID), which allows the database engine to locate the persistent representation of the object. In addition, the ODBMS engine manages the dynamic behavior (methods, polymorphism, inheritance) of the objects delivered to the user by setting up the proper function table. Unfortunately when using existing implementations, this mechanism is implementation specific, and does not allow reference to objects outside the current database engine.

To overcome such limitations a generic persistent model was developed to allow the following actions:
- Select the correct storage engine to access the object with its desired shape.
- Locate the object on the persistent storage medium.
- Read the object data and the object references.
- Handle the object's dynamic behavior by setting up the proper virtual function table through the invocation of the constructor.

Our design assumes that most database technologies are based on files or logical files. Internally these files are partitioned into containers ("Root trees" or "Root directories" for ROOT I/O, tables in relational database technologies) and objects populating these containers (see Figure 6). Objects within a container are addressed using a record identifier.

Using these back-end persistency technologies, a generic token was designed, which supports the above functionality. The database technology identifier, the object type, the database file, the container/table and the object identifier within the container form a universal object address, which allows data items to be addressed in nearly any known technology. This information is stored in the token, which fully identifies an object and forms a relocatable object address.

The full data content of such a token can be relocated between processes and individual users e.g. to communicate the identity of a physics event by electronic mail. Since the token fully identifies the object within any persistent medium, the token is equivalent to the object itself and hence allows to easily load objects on demand using a smart pointer mechanism. One implementation of such a mechanism is the POOL data cache as described in section 2.1.

To store relationships between objects in a logical database, this address is split, to minimize persistent storage overhead. As illustrated in Table 1 for various technology choices, the object type, database technology, the database name, the container name are stored in a separate lookup table, identified by a primary key, the link identifier. Hence, a persistent reference only contains the key to this look-up table and the record id of the object in the corresponding container. An instance of this look-up table is stored in every logical database and allows resolving all references from objects within the same logical database to any other object in- or outside this database.

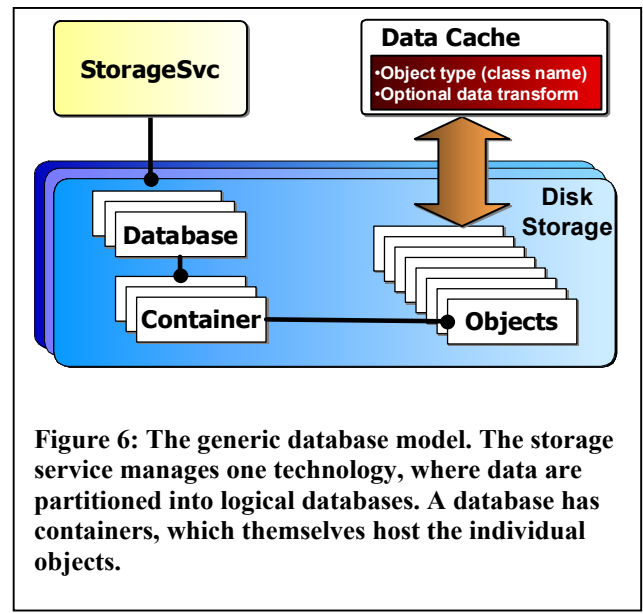

**Figure 6: The generic database model. The storage service manages one technology, where data are partitioned into logical databases. A database has containers, which themselves host the individual objects.**

| Technology Type | ROOT-Keyed | ROOT-Tree | RDMS | ORACLE |
|---|---|---|---|---|
| File / Database identifier | ROOT File name | ROOT File name | Database name | User login/table space |
| Container Identifier | Directory | Tree/Branch | Table name | Table/View name |
| Object identifier | Unique key | Record Number | Primary Key | Primary Key |

**Table 1 Layout of the universal container address in the link table. The role of the database name and the container name depends on the persistent technology.**

This local object resolution approach ensures, that data within individual logical databases are self-consistent. The local uniqueness of the objects is extended to global uniqueness by granting a unique naming of the individual database identifiers and typical problems of commercial solutions, trying to define object uniqueness at the level of small object identifiers is circumvented. The database identifiers are uniquely represented by a GUID, where a GUID can be transformed into the physical file name using the file catalog component.

When writing, this look-up table is automatically updated whenever a new link that is to be made persistent occurs in the object model.

If the token represents an object in an object database, an object association can be directly represented by the duple consisting of the record identifier and the object identifier. Using such a shortcut, the additional lookup in the indirection table may be omitted. Otherwise all information required locating the database file, the container/table and the corresponding record must be determined using a lookup.

## 3. EXPERIENCE WITH VARIOUS PERSISTENCY SOLUTIONS

The model described above has been implemented for several back-end technologies. In ATLAS, CMS and LHCb the POOL data storage mechanism using ROOT I/O as a backend solution is currently the preferred solution to write event data. Another implementation for a persistent backend implementation using relation database technology is under development.

The system has been tested so far on a rather small scale and performs well. A detailed analysis of the additional cost with respect to CPU and persistent storage is planned the before the deployment on a larger scale and the integration into the experiment frameworks. The additional overhead to implement the object reference mechanism is with eight Bytes per association rather small.